\documentclass[sigconf]{acmart}

\usepackage{natbib}
\usepackage{subfigure}
\usepackage{svg}
\usepackage{multirow}
\usepackage{balance}

\AtBeginDocument{%
  \providecommand\BibTeX{{%
    \normalfont B\kern-0.5em{\scshape i\kern-0.25em b}\kern-0.8em\TeX}}}

\setcopyright{acmcopyright}
\copyrightyear{2023}
\acmYear{2023}
\acmDOI{XXXXXXX.XXXXXXX}

\acmConference[CIKM '23]{32nd ACM International Conference on Information and Knowledge Management}{October 21--25,
  2023}{Birmingham, UK}
\acmPrice{15.00}
\acmISBN{978-1-4503-XXXX-X/23/10}

\copyrightyear{2023}
\acmYear{2023}
\setcopyright{acmlicensed}\acmConference[To appear in the proceedings of CIKM '23]{To appear in the proceedings of the 32nd ACM International Conference on Information and Knowledge Management}{October 21--25, 2023}{Birmingham, UK}
\acmBooktitle{Proceedings of the 32nd ACM International Conference on Information and Knowledge Management (CIKM '23), October 21--25, 2023, Birmingham, United Kingdom}
\acmPrice{15.00}
\acmDOI{10.1145/3583780.3615256}
\acmISBN{}

\begin{document}

\title{Findability: A Novel Measure of Information Accessibility}

\author{Aman Sinha}
\affiliation{%
  \institution{Indian Institute of Science Education and Research, Kolkata, India}
  \city{}
  \country{}
}
\email{as18ms065@iiserkol.ac.in}
\author{Priyanshu Raj Mall}
\affiliation{%
  \institution{Indian Institute of Science Education and Research, Kolkata, India}
  \city{}
  \country{}
}
\email{prm18ms118@iiserkol.ac.in}
\author{Dwaipayan Roy}
\affiliation{%
  \institution{Indian Institute of Science Education and Research, Kolkata, India}
  \city{}
  \country{}
}
\email{dwaipayan.roy@iiserkol.ac.in}
%
\renewcommand{\shortauthors}{Aman Sinha, Priyanshu Raj Mall, \& Dwaipayan Roy}

\begin{abstract}
The overwhelming volume of data generated and indexed by search engines poses a significant challenge in retrieving documents from the index efficiently and effectively. Even with a well-crafted query, several relevant documents often get buried among a multitude of competing documents, resulting in reduced accessibility or ``findability" of the desired document. Consequently, it is crucial to develop a robust methodology for assessing this dimension of Information Retrieval (IR) system performance.
While previous studies have focused on measuring document accessibility disregarding user queries and document relevance, there exists no metric to quantify the findability of a document within a given IR system without resorting to manual labor. This paper aims to address this gap by defining and deriving a metric to evaluate the findability of documents as perceived by end-users. Through experiments, we demonstrate the varying impact of different retrieval models and collections on the findability of documents. Furthermore, we establish the findability measure as an independent metric distinct from retrievability, an accessibility measure introduced in prior literature.
\end{abstract}



\maketitle

\section{Introduction}

In the present digital era, where information is abundant and easily available, the challenge lies not in obtaining information but in finding the right information when we need it. 
Whether searching for a specific record in a vast database, locating a particular product on an e-commerce website, or identifying important content on the internet, finding relevant information has become a crucial aspect of our daily lives.
Active research in the domain of information retrieval continues to evolve and address the various challenges emerging from the complex information need of the users.
One of the most explored areas of research is in exploratory search where the purpose is to discover new information, gain a deeper understanding of a subject, or explore different perspectives; in this scenario, the users are likely to be engaged in an open-ended search process to discover and learn about a particular topic or subject~\cite{exploratorySearch_white}.
In these searches, users have a general idea of what they are interested in but may not have specific details or keywords in mind. 
However, various biases in the retrieval systems may also play an adverse role in the selection of documents among the top positions after a retrieval~\cite{wilkie_2014, samar_ijdl}, making a significant amount of documents non-accessible to the user~\cite{retrievability}.

On the other hand, known-item searches occur when a user is looking for a specific item or a particular piece of information. 
In this type of search, the user already has a clear idea of what they are searching for and often possesses some specific details about the item (text document in our context), such as a title, author, or specific keywords. 
The goal is to \emph{find} an exact document quickly and accurately. 
However, finding a particular document can be challenging due to factors such as the large volume of available information, document variability, and biases in retrieval models.

In this paper, we propose a novel measurement scheme to quantify the overall findability of items in a document collection.
Considering a collection of entities or items, if we assume that each entity will be relevant to a subset of all possible queries, the \emph{findability} of an entity is the quantification of the ease with which the item can be found by the user by issuing queries from that subset only.
To evaluate the practicality and effectiveness of the findability measure, we conduct an empirical study on benchmark retrieval datasets.
By conducting experiments with various retrieval models and datasets of varying sizes, we observe the diverse effect and implications on the findability within an IR system.


\section{Background and related work}\label{sec:rel-work}

Research on the accessibility of documents in the domain of IR has focused on various aspects, such as crawlability~\cite{crawlability}, findability~\cite{amb-findability}, discoverability~\cite{discoverability}, retrievability~\cite{retrievability} and navigability~\cite{navigability}.
%
%
%
The accessibility of information within a collection, as addressed in existing literature, encompasses two distinct perspectives: 
from the system's standpoint~\cite{retrievability}, and 
from the user's standpoint~\cite{amb-findability,pagehunt}. 
Retrievability~\cite{retrievability} measures gauge the ease with which a document can be retrieved utilizing a specific IR system. In contrast, the concept of findability~\cite{amb-findability} seeks to assess the ease with which a user, utilizing the IR system, can locate a particular document\footnote{While the FAIR principle~\cite{wilkinson2016FAIR} distinguishes between accessibility and findability, in this study, we use the term findability to signify a method for gaining access to the documents within a collection.}.

The retrievability measure focuses exclusively on the system-oriented aspect of document retrieval, neglecting user intent during a search. Further, the retrievability measure provides only a general approximation of the likelihood of a document being retrieved, regardless of which query is posed to the IR system. This approximation merely attempts to gauge the accessibility of documents within the collection facilitated by the IR system, disregarding the user's perspective. When users interact with a search engine, they express their specific information need by submitting a query, reflecting their intent and purpose for utilizing the IR system. Therefore, to accurately estimate document accessibility within the collection, it is crucial to account for both the user and their query together with the IR system during the search process.

The ability to access relevant documents in response to a user's query lies at the heart of document findability from the user's standpoint. For instance, presenting a document related to "Java" in response to a user's search for "Python" does not contribute to the findability of "Java" because the document lacks relevance to the query posed. In this context, the findability of a document refers to its capacity to be located solely for queries whose intent is satisfied or addressed by that particular document. In other words, a document is considered \emph{found} when the user is satisfied to encounter it in the search results as per the query they entered into the IR system to find that document.
A similar approach was followed in~\cite{pagehunt} where the authors introduced \emph{Page Hunt}, a game specifically designed to collect web search log data. The game involves presenting participants with webpages and tasking them with finding them using the provided search interface. 

Previous studies~\cite{azzopardi2008accessibility,retrievability} have introduced retrievability measures to estimate the ease of document retrieval using search engines. Additionally, other researchers~\cite{chi2000scent,pandit2007navigationaided,zhou2007mnav} have explored something like findability from the perspective of browsing and navigability, evaluating how easily users can navigate websites. Authors of~\cite{azzopardi2013towards} say that successful validation of findability measures could enable the development of tools to assist Information Architects in analyzing websites, offering insights into the findability and utilization of content, as well as identifying features (such as terms, links, etc.) that contribute to the ease or difficulty of locating specific pages.

\section{Findability - a measure of accessibility}\label{sec:findability}

Consider an IR system that employs model $R$ to retrieve relevant documents in response to user queries from a document collection $D$. 
Let $Q_d$ represent the set of all possible queries for which document $d$ (where $d \in D$) is deemed relevant. 
We refer to these queries as ``relevant queries'' specifically for that particular document.
The \textit{findability} of a document is then defined as the expectation of the likelihood of a user finding that document for every query $q$ ($q \in Q_d$).
Mathematically, the findability measure is formulated as:
\begin{equation}\label{eq:findability}
    f(d) = \frac{1}{|Q_d|} \sum_{q \in Q_d} \xi (p_{dq} , c)
\end{equation}
%
In Equation~\ref{eq:findability}, $p_{dq}$ is the rank of document $d$ in the search result against query $q$.
The function $\xi (p_{dq}, c)$ is a generalized convenience function that captures users' willingness to explore the search results up to rank $p_{dq}$, while $c$ denotes the threshold rank at which users cease examining the ranked list.

The function $\xi (p_{dq}, c)$ is subject to two boundary constraints. The first constraint ensures that when document $d$ appears at the top rank ($p_{dq} = 1$), it represents the most convenient and optimal scenario for the user; $\xi (1, c)$ is set to 1 to reflect this favorable situation.
On the other hand, users typically do not continue indefinitely exploring the ranked list of search results; they stop investigating (denoted by $c$ in Equation~\ref{eq:findability}) at some point. 
In the worst-case scenario, if a document appears in the search results after the user has stopped investigating, it means the user does not find that document.
Thus, for ranks $p_{dq}$ greater than $c$, we set $\xi (p_{dq}, c) = 0$. 
With these two constraints, the function $\xi(.)$ is bounded within the range of $[0,1]$, making it a suitable measure for interpretation.

To define the convenience function $\xi(.)$, we employ the concept of Click Through Rate (CTR): the net percentage of clicks that a document at a certain rank gets out of total clicks by users to open a search result document in the ranked list.
The CTR on a search engine could be taken as a representation of the user tendency or user effort it takes to investigate a certain rank in the results. 
Notably, analyses conducted by Semrush Inc. and Backlinko\footnote{https://backlinko.com/google-ctr-stats} based on 4 million Google search results offer valuable insights into CTR for top ranks in the context of web searches. 
These findings indicate that a mere $0.63\%$ users click on results beyond rank position 10, suggesting that a majority of users discontinue exploring the ranked list after a certain rank threshold.
The observations from CTR of users lead us to propose the following two forms for the convenience function $\xi (p_{dq}, c)$ in the context of findability measure:

\begin{itemize}
    \item \textbf{Exponential decay of Convenience:}
    Considering an exponential decay with a decay rate of approximately one-third, the convenience function can be defined as:
%
    %
    \begin{equation}\label{eq:xi1}
        \xi (p_{dq}, c) =
        \begin{cases}
            e^{-(p_{dq}-1)/3} & \text{if } p_{dq} \leq c \\
            0 & \text{if } p_{dq} > c
        \end{cases}
    \end{equation}
    \item \textbf{Inverse law of Convenience:}
    An alternative approach to incorporate the decaying effect is by considering the inverse of the document rank, which can be expressed as follows:
    \begin{equation}\label{eq:xi2}
        \xi (p_{dq}, c) =
        \begin{cases}
            \frac{1}{p_{dq}} & \text{if } p_{dq} \leq c \\
            0 & \text{if } p_{dq} > c
        \end{cases}
    \end{equation}
\end{itemize}

\subsection{Estimating Document Findability} \label{subsec:est-find}
In order to estimate \textit{findability} scores of documents in an operational setting, one crucial requirement is the creation of a relevant query set for every document. 
Ideally, we would require a comprehensive list of all the queries for which a document could be considered relevant. 
However, generating such an exhaustive list manually would be an impractically labor-intensive task, even for a moderately-sized collection of documents:
human experts need to read the documents and submit search queries that are deemed relevant to the respective documents.
Avoiding the involvement of human efforts, a known-item query generation strategy can be employed as a proxy for automatically generating a smaller but representative sample of relevant queries for each document.

\begin{table}[]
    \centering
    \caption{Statistics of datasets used for experimentation.}
    \label{tab:Dataset Statistics}
    \resizebox{1.0\columnwidth}{!}{
    \begin{tabular}{lrcrrr}
    \hline
        \textbf{Dataset} & \textbf{\# documents} & \textbf{Collection type}  & \textbf{\# terms} & \textbf{\# queries} \\ \hline
        \textbf{TREC Robust} & 528,155 & News & 1,502,031 & 10,230,070 \\
        \textbf{WT10g} & 1,692,096 & Web & 9,674,707 & 26,041,327 \\
        \textbf{MS MARCO passage} & 8,841,823 & Web excerpts  & 1,410,558 & 19,839,452 \\ \hline
    \end{tabular}
    }
\end{table}

\subsection{Mean Findability and Findability Bias}
When evaluating document findability within a fixed collection, two aspects of access provided by different retrieval models can be assessed. Firstly, the \emph{mean of findability} scores for all documents in the collection reflects the overall effectiveness of the retrieval model in retrieving the correct document at the top. This provides a measure of the aggregate performance of the retrieval model in delivering relevant documents.
Secondly, the Gini coefficient~\cite{gini1936measure} can be employed to quantify the disparity of access imposed by the retrieval model across the collection. In this context, the Gini coefficient represents the imbalance of access among the documents within the collection itself. It captures the extent to which certain documents are favored over others in terms of findability.
By considering both the mean findability, denoted as $\left\langle f \right\rangle$, and the findability bias, represented by the Gini coefficient $G$, a comprehensive assessment of document findability offered by a retrieval model can be obtained. This dual approach provides a holistic understanding of how effectively and fairly the retrieval model enables access to the documents in the collection.

The applicability of Gini coefficient as a measure of inequality in the context of accessibility in IR has been employed in earlier studies~\cite{retrievability,inequality_measures}. 
This measure, borrowed from economics and social sciences, provides a quantitative way to assess the level of inequality in access to information. 
Lorenz curve~\cite{lorenz1905methods,lorenz-gini}, the graphical representation of Gini coefficient is used to visualize the deviation of wealth distribution from equality.
In the context of findability, the Gini coefficient can be computed as follows:
\begin{equation}
    G = \frac{\sum_{i=1}^{N} (2i-N-1)\cdot f(d_i)}{N \sum_{j=1}^{N} f(d_j)}
\end{equation}
where $f(d_i)$ and $f(d_j)$ are $i^{ith}$ and $j^{th}$ documents when documents are ordered in ascending order by their findability scores; $N$ is the total number of documents in the collection. 
Gini coefficient $G$ ranges from 0 to 1, where $G = 0$ means no bias (i.e., an equal distribution which implies $f(d)$ is equal for all documents) and $G = 1$ indicates maximum bias (i.e., all documents have $f(d) = 0$ except one document) - implying that only one document consistently appears at the top ranks for its relevant queries, while the rest of the documents are never found within the top ranks and remain hidden among other documents (that take up the top $c$ positions).

\begin{table}[htp]
    \caption{Gini coefficient $G$ and Mean Findability $\left\langle f \right\rangle$ for Findability $f(d)$}
    \label{tab:exp1.1}
\begin{tabular}{llccc}
\toprule
        & & \textbf{Robust04} & \textbf{WT10g}  & \textbf{MS MARCO} \\ \hline
\multirow{2}{*}{\textbf{LM-Dir}} 
        & $G$ & 0.1587   & 0.2847 & 0.3774  \\
        & $\left\langle f \right\rangle$ & 0.6327   & 0.5209 & 0.5173  \\ 
        \hline
\multirow{2}{*}{\textbf{BM25}} 
        & $G$ & 0.1456   & 0.2503 & 0.3116  \\
        & $\left\langle f \right\rangle$ & 0.6640   & 0.5985 & 0.5895  \\ 
        \hline
\multirow{2}{*}{\textbf{DFR-PL2}} 
        & $G$ & 0.1424   & 0.2497 & 0.3007  \\
        & $\left\langle f \right\rangle$ & 0.6672   & 0.6133 & 0.5888  \\ 
        \bottomrule
\end{tabular}
\end{table}
\begin{table}[h!]
    \caption{Correlation between Findability $f(d)$ and Retrievability $r(d)$ when both $f(d)$ and $r(d)$ are estimated using query sets generated from their respective datasets (Self query set) as well as the known-item search query set (Known-item query), where all are statistically significant correlation with $p < 0.05$.}
    \label{tab:exp2.1}
    \centering
    \resizebox{0.99\columnwidth}{!}{
    \begin{tabular}{l|rr|rr}
    \hline
    & \multicolumn{2}{c|}{Self query set} & \multicolumn{2}{c}{Known-item query set} \\ \hline
    & \textbf{Pearson's r} & \textbf{Kendall's $\boldsymbol{\tau}$} & \textbf{Pearson's r} & \textbf{Kendall's $\boldsymbol{\tau}$} \\ \hline
    \textbf{Robust04} & -0.0944 & -0.0518  & -0.1292 & -0.1053  \\ \hline
    \textbf{WT10g}    & -0.0088 & 0.0084 & -0.0256 & -0.0287  \\ \hline
    \textbf{MS MARCO}  & 0.0115 & 0.0307 & 0.0388 & 0.0269   \\ \hline
    \end{tabular}
    }
\end{table}

\section{Empirical analysis}\label{sec:exp}

In this section, we present an experiment that showcases a practical use-case scenario for the findability measure.
We evaluate the Mean Findability and Findability Bias of three standard retrieval models across three different benchmark collections to determine which retrieval model offers the best accessibility in diverse collection types.
This evaluation allows us to assess the effectiveness of each retrieval model in retrieving the correct documents at the top ranks and the degree of bias in document access within each collection.
Further, we investigate the relationship between findability and retrievability~\cite{azzopardi2008accessibility}. By comparing findability scores with retrievability scores, we uncover that findability scores are independent and distinct from retrievability scores. 
%
%
%
%

\subsection{Datasets and Retrieval Models}

We evaluate the proposed findability metric using three benchmark datasets: TREC Robust, WT10g, and MS-Marco passage. 
The statistics of the datasets are presented in Table~\ref{tab:Dataset Statistics}.
These datasets are commonly used in information retrieval research and provide diverse document collections for our analysis.
In this study, we investigate the findability of documents using three different retrieval models, particularly BM25~\cite{Robertson1994Okapi, sparkjones_probabilistic}, 
LM-Dir~\cite{zhai_lafferty}, 
and DFR-PL2~\cite{dfr}.

\begin{figure*}[t!]
    \centering
    \subfigure{\includegraphics[width=0.33\textwidth]{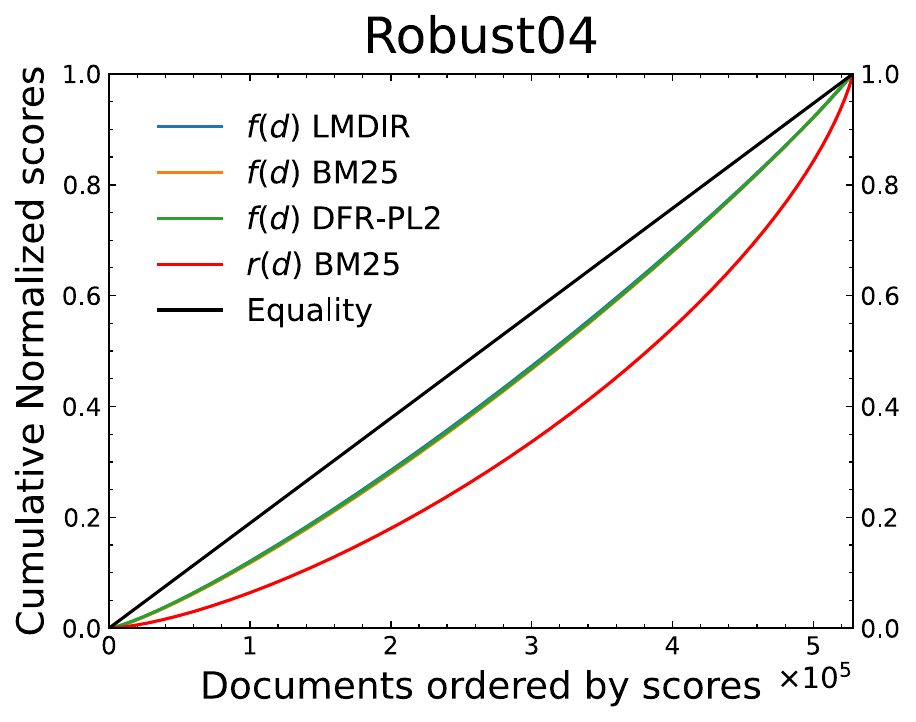}}
    \subfigure{\includegraphics[width=0.33\textwidth]{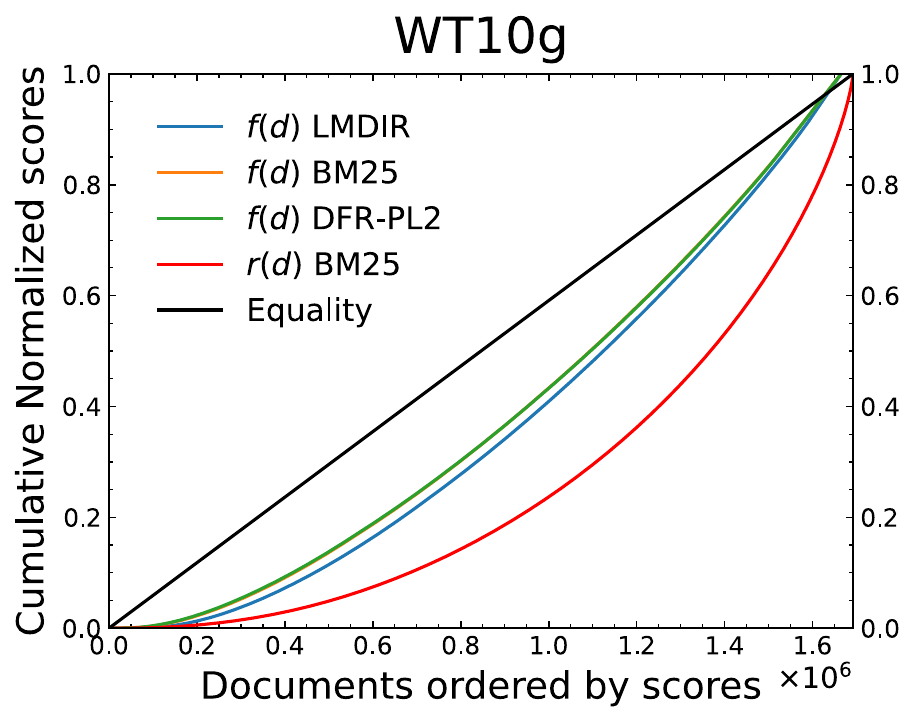}}
    \subfigure{\includegraphics[width=0.33\textwidth]{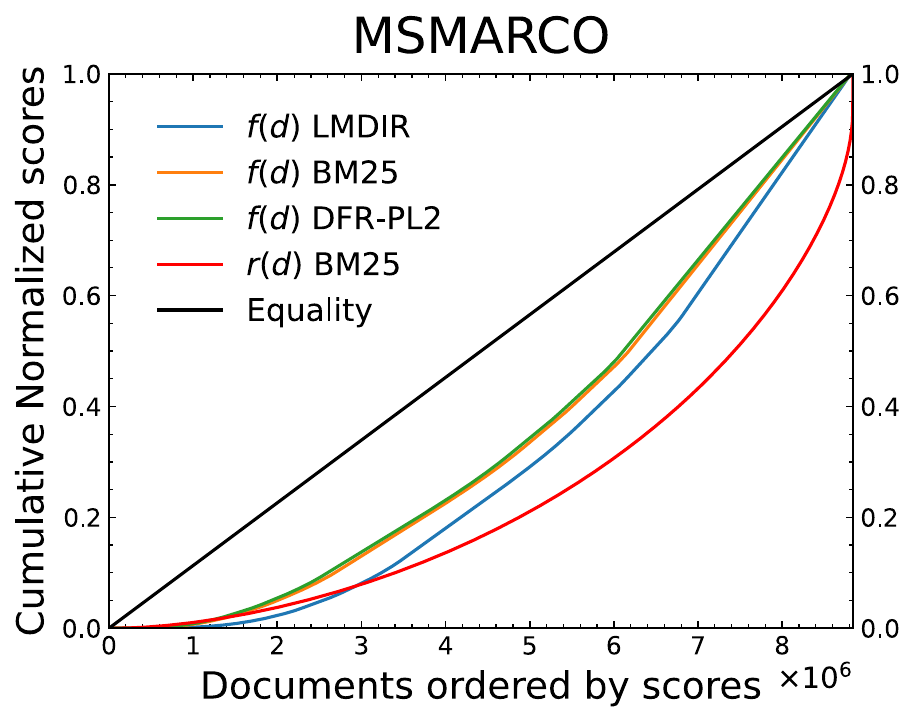}}
    \caption{Lorenz Curves for three retrieval models for each of the three collections. 
    Note that, some of the curves are closely situated, potentially making them indistinguishable in the plots, particularly for the Robust04 and WT10g collections.
    To highlight the contrast between retrievability with findability, the scaled curves for retrievability are also presented in red.
    }
    \label{fig:lorenz_curves}
\end{figure*}

\subsection{Query Generation Method}

As discussed in Section~\ref{subsec:est-find}, the estimation of findability requires a set of relevant queries for each document in the collection. 
In this study, we employ a known-item query generation method, which was introduced in a previous work~\cite{simulated_query}. 
This method has demonstrated its effectiveness in generating query scores that are comparable to manually generated known-item queries. 
Following the \textit{popular+discriminative} selection strategy method~\cite{simulated_query}, we generate a set of known-item search queries with the average query length $k$ and mixing parameter $\lambda$ (as defined in~\cite{simulated_query}) set to $4$ and $0$ respectively. 


To ensure a comprehensive assessment, the generated number of relevant queries is set to 10\% of the total number of distinct terms in the document with an upper cap of $50$ to maintain computational tractability.

\subsection{Experimentation}

The findability measure, defined in Equation~\ref{eq:findability}, includes a parameter $c$ that represents the maximum rank tolerance of users.
While the optimal value of $c$ can vary depending on a specific task and user preferences, we have chosen a value of $c=100$ for this particular study\footnote{We experimented with $c$ from 10 to 100 varied in steps of 10; considering the space limitation, we are reporting the results for $c=100$ in this paper.}.
Based on our initial study of CTR data, we found that the inverse law of convenience provides a better fit. Therefore, we have opted to utilize the inverse law form (Equation~\ref{eq:xi2}) for estimating the findability scores.

We report the Gini coefficient $G$ and the mean findability $\left\langle f \right\rangle$ of the collections in Table~\ref{tab:exp1.1}.
From the table, we can observe that as the Gini coefficient increases, there is a noticeable decrease in mean findability. 
This implies that across the three examined retrieval models, improving the overall findability of documents results in a concurrent decrease in findability bias.
Consequently, retrieval models that increase the findability of documents tend to improve accessibility to the collection as a whole by enhancing the findability of the majority of documents.
The inequality is graphically presented as Lorenz curve in Figure~\ref{fig:lorenz_curves}. In the figure ~\ref{fig:lorenz_curves}, the curve for retrievability employing BM25 model is also presented in red to showcase the disparity with the findability bias.

Moreover, it is observed that mean findability decreases and findability bias increases with collection size. This aligns with our intuition that a larger number of documents results in heightened competition among them for higher rankings, thereby diminishing findability. Additionally, it seems that the bias imposed by the retrieval model on the collection becomes more pronounced as the collection size increase.

Among the three retrieval models, LM-Dir yields lower findability for the collection's documents compared to the other two models. While BM25 and PL2 exhibit similar mean findability, the Gini coefficient for PL2 is reported to be lower indicating that PL2 performs better overall in terms of the findability aspect of accessibility-based model performance.



\subsection{Comparing with Retrievability}

The distinction between retrievability and findability may initially appear subtle. Still, it is essential to acknowledge the substantial conceptual distinctions between them that emerge when considering user behavior and user-centric factors.
To clarify the distinction, we perform a retrievability analysis using BM25 retrieval model for all three collections, utilizing a standard query set commonly used for computing retrievability scores, as mentioned in the original work~\cite{retrievability}. 
Further, we compute Pearson's and Kendall's rank correlation coefficient. As presented in Table~\ref{tab:exp2.1}, the obtained correlations reveal an almost negligible association between the two measures. 
Even when utilizing the same queries for retrievability analysis as those employed for findability evaluation (known-item queries), Table \ref{tab:exp2.1}'s third column reveals a persistent lack of relationship between findability and retrievability. 
These correlation results serve to establish findability as an independent measure of accessibility that was not previously encompassed by the retrievability measure.

The findability score provides a uniform interpretation and a constant range, akin to a coefficient, unlike retrievability which encounters comparability challenges across diverse studies due to variations in query sets.
Moreover, findability measure is well-suited for analyzing the findability of individual documents, whereas a single retrievability score alone may not adequately represent retrievability of a document without additional context or information.

\section{Conclusion and future work}\label{sec:conclu}

This paper introduces a novel metric called `findability' for measuring document accessibility. 
Our study demonstrates that all three retrieval models exhibit comparable behavior in regard to findability.
Furthermore, we compare findability with retrievability, another existing metric for document accessibility. Future research will investigate the use of the findability measure in fine-tuning IR system parameters in situations where relevance judgments are unavailable. Additionally, exploring the correlation between improved overall findability and enhanced user experience is an area of interest for further investigation.

\noindent
\emph{\textbf{Acknowledgement:}}
We would like to thank the anonymous reviewer for their valuable and encouraging feedback.

\bibliographystyle{ACM-Reference-Format}
\balance
\bibliography{arxiv}

\end{document}